\documentclass[aip,preprint,superscriptaddress,floatfix]{revtex4-1}

\usepackage{mathptmx}
\usepackage[scaled=0.92]{helvet}
\usepackage[utf8]{inputenc}
\usepackage{graphicx}
\usepackage[dvipsnames]{xcolor}
\usepackage[pdftex]{hyperref}
\hypersetup{colorlinks,citecolor=blue}
\usepackage[T1]{fontenc}

\usepackage[textwidth=1.5cm]{todonotes}
\usepackage{textcomp}
\usepackage{ifsym}
\usepackage{mathrsfs}
\usepackage{sidecap}
\usepackage[
,textwidth=17.5cm
,textheight=23.5cm
,verbose
,dvips
]{geometry}
\usepackage{xspace}

\newcommand{\celsius}{$^{\circ}$C\xspace}

\begin{document}

\title{Dependence of the Mn sticking coefficient on Ga-rich, N-rich, and Ga/N-flux-free conditions in GaN grown by plasma-assisted molecular beam epitaxy}

\author{YongJin~Cho}
\email[Author to whom correspondence should be addressed: ]{cho@pdi-berlin.de}
\affiliation{School of Electrical and Computer Engineering, Cornell University, Ithaca, New York 14853, USA}
\affiliation{Paul-Drude-Institut für Festkörperelektronik (PDI), Leibniz-Institut im Forschungsverbund Berlin e.V., Hausvogteiplatz 5--7, 10117 Berlin, Germany}
\author{Changkai~Yu}
\affiliation{Department of Materials Science and Engineering, Cornell University, Ithaca, New York 14853, USA}
\author{Huili~Grace~Xing}
\affiliation{School of Electrical and Computer Engineering, Cornell University, Ithaca, New York 14853, USA}
\affiliation{Department of Materials Science and Engineering, Cornell University, Ithaca, New York 14853, USA}
\affiliation{Kavli Institute at Cornell for Nanoscale Science, Cornell University, Ithaca, New York 14853, USA}
\author{Debdeep~Jena}
\affiliation{School of Electrical and Computer Engineering, Cornell University, Ithaca, New York 14853, USA}
\affiliation{Department of Materials Science and Engineering, Cornell University, Ithaca, New York 14853, USA}
\affiliation{Kavli Institute at Cornell for Nanoscale Science, Cornell University, Ithaca, New York 14853, USA}

\begin{abstract}
This brief report examines the influence of Ga/N flux conditions on Mn incorporation in GaN. Mn-doped GaN layers were grown at 680\celsius by molecular beam epitaxy on a Ga-polar GaN(0001) template substrate under Ga-rich, N-rich, and no-flux conditions (i.e., Mn $\delta$ doping). Mn incorporation was highest under N-rich condition, lowest under Ga-rich condition, and intermediate in the absence of Ga and N fluxes. For the growth conditions examined in this study, the corresponding Mn sticking coefficients, relative to that of the N-rich condition, were determined to be 0.31 for no-flux growth and 0.01 for the Ga-rich growth.

\end{abstract}


\maketitle
\section{Introduction}
Nitride semiconductors are among the most commercially important semiconductor materials after Si. They have been extensively employed in a wide range of electronic and optoelectronic devices owing to their wide bandgap tunability and both \textit{n}- and \textit{p}-type dopability. In recent years, growing interest has focused on incorporating new functionalities into nitrides, including ferroelectricity and superconductivity.\cite{jena2019new} Mn, a transition-metal ion with a half-filled \textit{d} orbital, provides large magnetic moments when incorporated into a semiconductor host,\cite{furdyna1988diluted,zhou2007ferromagnetic,cho2008magnetic} enabling potential spintronic applications. Both ferromagnetic GaMnN (Refs.~\onlinecite{kondo2002molecular,sonoda2002molecular,sawicki2012origin,gas2021improved}) and ferrimagnetic Mn$_{4}$N (Refs.~\onlinecite{dhar2005ferrimagnetic,zhang2020magnetic,zhang2021epitaxial}) have been explored as approaches to introduce magnetic functionality into the nitride material platform. On the other hand, Mn can also be used to achieve semi-insulating GaN over a wider temperature range than Fe- or C-doped GaN, owing to its deeper acceptor activation energy.\cite{bockowski2018doping} 

Although N-rich growth conditions are commonly employed to enhance Mn incorporation in GaN,\cite{haider2003ga,stefanowicz2013phase,sawicki2012origin} the influence of Ga/N flux conditions on Mn incorporation during molecular beam epitaxy (MBE) growth has not been systematically investigated. This is despite the fact that, in plasma-assisted MBE, N-rich conditions are generally unfavorable for GaN growth due to the tendency toward three-dimensional growth modes.\cite{heying2000control} In this work, we systematically examine the effect of Ga/N flux conditions on Mn incorporation and determine the corresponding relative Mn sticking coefficients.

\section{Experimental}

A GaN:Mn/GaN multilayer sample was grown on a Ga-polar GaN(0001)/sapphire template in a Veeco GENxplor MBE chamber equipped with standard dual-heater effusion cells for Ga and Mn and a radio frequency plasma source for active N species. The base pressure of the growth chamber was $\approx$10$^{-10}$~Torr under idle conditions, and $\approx$2$\times$10$^{-5}$~Torr during growths due to N$_2$ gas. An rf plasma power of 200~W and a nitrogen flow rate of 1.9~sccm were used, which were found to lead to a growth rate of 6.1~nm~min$^{-1}$ for \textit{c}-plane GaN under Ga-rich conditions, i.e., N flux $f_{N}=6.1$~nm~min$^{-1}$. Two different Ga fluxes ($f_{Ga}$) of 8.0~nm~min$^{-1}$ and 4.8~nm~min$^{-1}$ were used for the Ga-rich ($f_{Ga}/f_{N}\approx1.3$) and the N-rich ($f_{N}/f_{Ga}\approx1.3$) conditions in the sample, respectively. All Mn-doped regions were grown using a fixed Mn beam-equivalent pressure of 1.6$\times 10^{-9}$~Torr in order to investigate the effect of the Ga/N flux ratio on Mn incorporation in GaN. The substrate temperature was estimated to be 680\celsius and was maintained during growth. The schematic of the sample structures is displayed in Fig.~\ref{figrheed}(a). A 225~nm GaN buffer layer was first grown on a GaN template in order to get a chemically clean GaN surface. Three different Mn-doped GaN regions were grown under Ga-rich, N-rich, and no flux conditions. The last condition, corresponding to Mn $\delta$-doping, was carried out by opening the Mn shutter, while the Ga and N shutters were closed. During this step, the N plasma source was ignited, and the chamber was filled with N$_{2}$ gas. Each Mn-doped region was sandwiched by GaN spacer layers grown under the same Ga-rich conditions. There were growth interruptions when the Ga-rich and N-rich conditions were switched, as indicated by the asterisks in Fig.~\ref{figrheed}(a). These interruptions were necessary to adjust the Ga cell temperature accordingly. The GaN surface was confirmed to be Ga-adlayer-free by in situ reflection high energy electron diffraction (RHEED) prior to the growth of the N-rich GaN:Mn and the Mn $\delta$-doped layers. The densities of Mn and the impurity elements H, C, O, and Si in the sample were measured by dynamic secondary ion mass spectrometry (SIMS). Mn was analyzed using magnetic-sector SIMS, while H, C, O, and Si were analyzed using quadrupole SIMS at two different EAG laboratories. These separate analyses resulted in a mismatch of approximately 25~nm between the two SIMS profiles, as indicated by the growth-interface peaks at $\approx$600~nm in Figs.~\ref{figsims}(a) and (b). Nevertheless, this discrepancy does not influence the evaluation of relative Mn incorporation under the different flux conditions. Both the as-grown MBE sample surface and the substrate surface exposed after the SIMS sputtering exhibit comparable root-mean-square roughness values of approximately 0.6~nm over a $2\times2$~$\mu$m$^{2}$ area [Figs.~\ref{figrheed}(b) and (c)].

\section{Results and Discussion}
\label{Res}

\begin{figure*}[t!]
\centering
\includegraphics*[width=13cm]{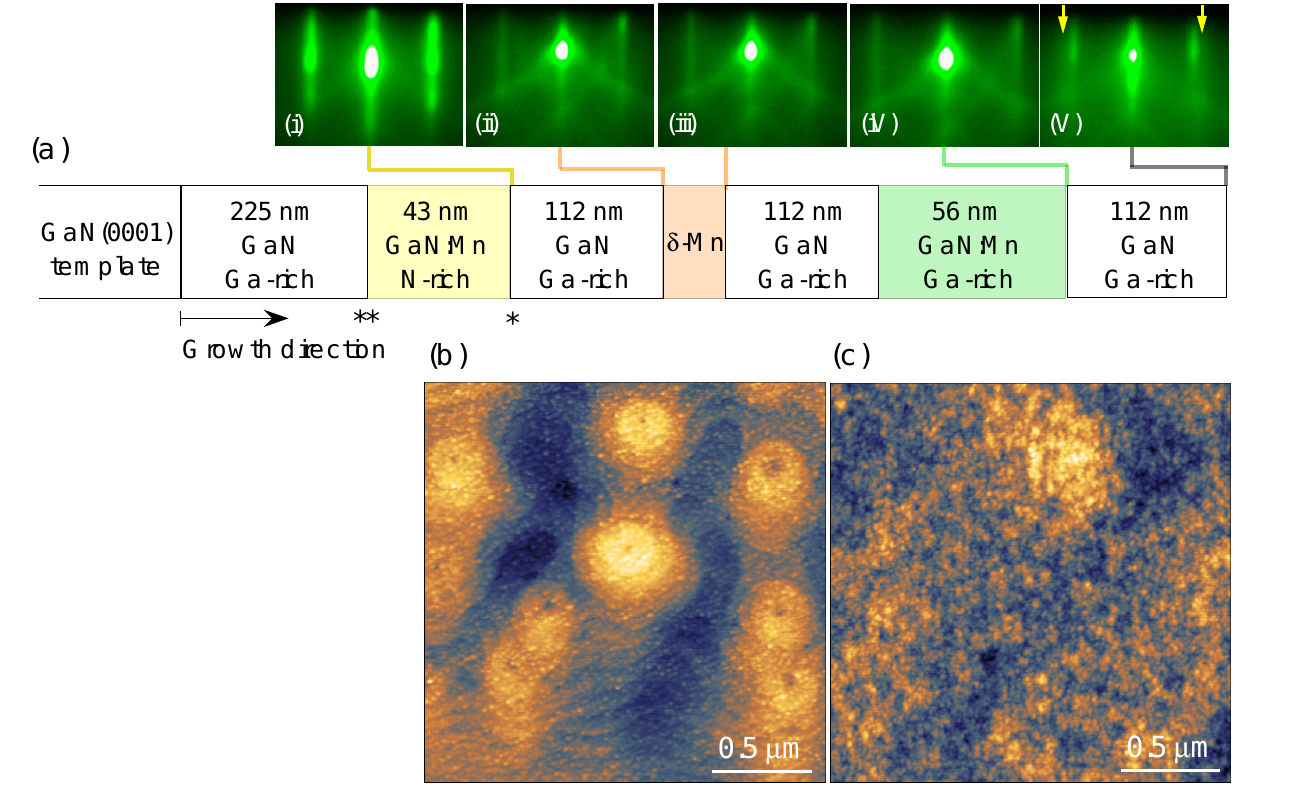} 
\caption{(a) Schematic layer structure of GaN:Mn/GaN multilayers grown on a GaN template at 680\celsius by MBE. The double and single asterisks indicate growth interruptions of 12~min. and 6~min., respectively. (i)--(v) RHEED patterns taken along the $<$11$\bar{2}$0$>$ azimuth at various stages of the growth, as indicated by the lines connected to (a). The two yellow arrows in (v) highlight the diffraction patterns associated with a contracted Ga bilayer on the GaN surface. [(b), (c)] AFM micrographs of the Mn-doped GaN multilayer structure, showing (b) the as-grown surface and (c) a sputtered substrate region after SIMS measurements. The RMS roughness over a $2\times2$~$\mu$m$^{2}$ area is 0.68~nm for (b) and 0.64~nm for (c).}
\label{figrheed}
\end{figure*}

\begin{figure*}[t!]
\centering
\includegraphics*[width=13cm]{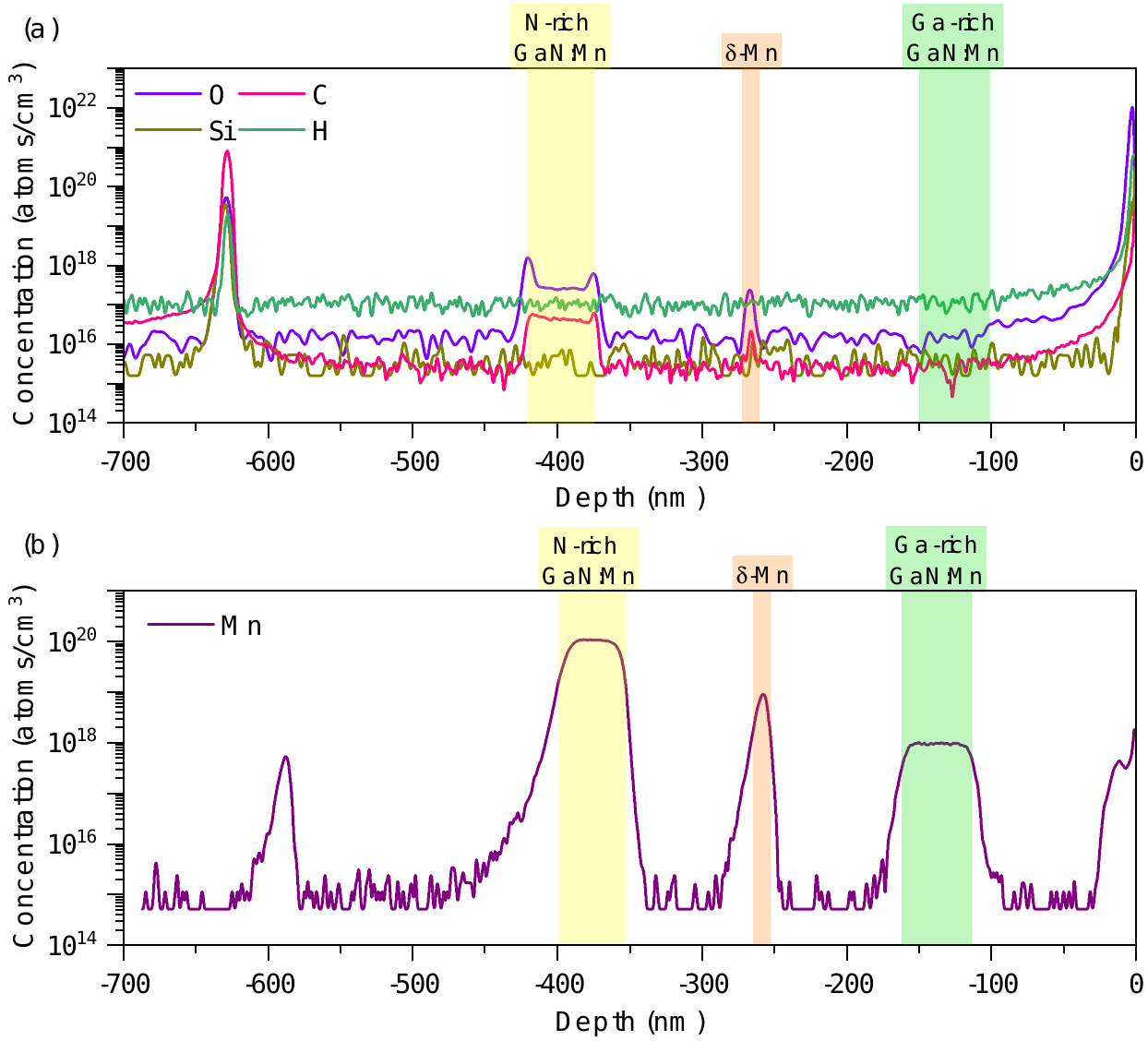} 
\caption{SIMS depth profiles of GaN:Mn/GaN multilayers grown on a GaN template at 680\celsius by MBE, showing (a) O, C, Si, and H and (b) Mn concentrations in the layers, with detection limits of $3\times10^{15}$~cm$^{-3}$ for Si and O, $1.5\times10^{15}$~cm$^{-3}$ for C, $1\times10^{17}$~cm$^{-3}$ for H, and $1\times10^{15}$~cm$^{-3}$ for Mn.}
\label{figsims}
\end{figure*}

Figures~\ref{figsims}(a) and (b) show SIMS depth profiles of the impurities O, C, Si, and H, and dopant Mn in the GaN:Mn/GaN multilayer sample. Before directly addressing Mn incorporation, we first briefly discuss the differences in impurity incorporation among the GaN:Mn regions. O and C concentrations under the N-rich and no-flux conditions exceed those under the Ga-rich conditions by more than one order of magnitude, indicating that a Ga-adlayer-free dry GaN surface is more prone to the incorporation of these impurities, particularly O. Such enhanced O and C incorporation under N-rich conditions is commonly observed in nitride films grown by plasma-assisted MBE.\cite{elsass2000effects,wang2021impact} The nearly constant impurity concentrations observed between the GaN buffer/barrier layers and the GaN:Mn layer grown under Ga-rich conditions indicate that the Mn flux does not affect those impurity incorporation within the experimental detection limits. A similar behavior is observed for Si and H, whose incorporation remains insensitive to the growth flux conditions, even under N-rich growth, indicating that surface kinetics are not the limiting factor.

We now examine how the Ga/N flux conditions influence Mn incorporation in GaN. The highest Mn concentration, approximately $\approx1\times10^{20}$~cm$^{-3}$, is obtained in GaN:Mn grown under N-rich conditions [Fig.~\ref{figsims}(b)]. During growth of this layer, the growth front becomes rough as a result of the reduced adatom diffusion length under N-rich conditions,\cite{heying2000control} as indicated by the modulated RHEED patterns [Fig.~\ref{figrheed}(i)]. The transmission patterns confirm that the wurtzite hexagonal crystal structure is maintained even at this high Mn incorporation level. The high Mn incorporation observed under N-rich conditions is unlikely to result from enhanced incorporation associated with the rough surface morphology; otherwise, the Mn density would be expected to increase gradually with the thickness of the N-rich-grown layer. An apparent Mn peak density approximately one order of magnitude lower is achieved under no Ga/N flux conditions, corresponding to the Mn $\delta$-doped region. The absence of changes in the RHEED patterns before [Fig.~\ref{figrheed}(ii)] and after [Fig.~\ref{figrheed}(iii)] Mn $\delta$-doping indicates that no structural degradation occurs as a result of the $\delta$-doping. Before proceeding to the Ga-rich results, it is useful to note that, although not central to the discussion, the apparent broadening of the Mn peak in the SIMS profile is largely governed by surface morphology. Because SIMS signals are collected from a relatively large sputtered area, the technique tends to broaden dopant profiles and underestimate the true volumetric density in spatially abrupt, narrow regions such as $\delta$-doped layers.\cite{gas2021improved} The Mn areal density ($\approx8.2\times10^{12}$~cm$^{-2}$, also see below) of the Mn $\delta$-doped region leads to a maximum possible volumetric density of $\approx3\times10^{20}$~cm$^{-3}$, assuming that the thickness of the Mn $\delta$-doped region is confined to a single GaN monolayer (ML). We now examine Mn incorporation under Ga-rich conditions, which exhibits a sharp decrease. The lowest Mn concentration, approximately $\approx1\times10^{18}$~cm$^{-3}$, is observed in GaN:Mn grown under Ga-rich conditions. The RHEED pattern after growth of the Ga-rich GaN:Mn layer [Fig.~\ref{figrheed}(iv)] is similar to that of a GaN layer grown under comparable Ga-rich conditions [Fig.~\ref{figrheed}(ii)]. Furthermore, the RHEED pattern [Fig.~\ref{figrheed}(v)] recorded below 300 °C after completion of the entire structure exhibits a pseudo-1 × 1 pattern characteristic of a Ga-polar GaN surface with a Ga bilayer,\cite{adelmann2003gallium,cho2019blue} indicating that no polarity inversion occurred as a result of the Mn doping. 

It is important to note that, under all the flux conditions, the widths of the Mn concentration plateaus closely correspond to the Mn shutter opening and closing times. This behavior indicates that any delayed Mn incorporation associated with Mn floating on the growth front is negligible, implying that Mn not incorporated during growth desorbs from the surface. Otherwise, the apparent thicknesses of the Ga-rich and $\delta$-doped Mn regions, as determined from the Mn SIMS profiles, would be considerably larger than the intended values. 

The areal densities of Mn in each Mn-doped region were determined by integrating the corresponding volume density profiles over depth. The integration limits were selected such that the Mn concentration reached the background signal. Based on the Mn shutter opening times (9~min for the N-rich and the Ga-rich GaN:Mn, and 34~s for the no-flux Mn $\delta$ doping) for each Mn-doped region, the resulting incorporated Mn fluxes were determined to be $7.8\times10^{11}$~atoms~cm$^{-2}$~s$^{-1}$, $2.4\times10^{11}$~atoms~cm$^{-2}$~s$^{-1}$, and $8.6\times10^{9}$~atoms~cm$^{-2}$~s$^{-1}$ for the N-rich, no-flux, and the Ga-rich conditions, respectively. These values yield relative Mn sticking coefficients, normalized to that under the N-rich condition, of 0.31 for the no-flux condition and 0.01 for the Ga-rich condition. A similarly significant difference in Mn incorporation between Ga-rich and N-rich conditions has been reported previously.\cite{kuroda2003strong} The observed dependence of the Mn sticking coefficient on the Ga/N flux conditions can be understood by considering that Mn occupies cation (Ga) sites in the GaN crystal.\cite{haider2003ga} Under N-rich growth conditions, Mn can readily access available cation sites. In contrast, under Ga-rich conditions, Mn \textit{competes} with Ga for incorporation at cation sites. This difference leads to different Mn desorption rates and sticking coefficients. A Ga-adlayer-free GaN surface, achieved in the absence of both Ga and N fluxes during the Mn $\delta$-doping, would result in a sticking coefficient that lies between the values observed under Ga-rich and N-rich conditions.


\section{Conclusions}
\label{Con}
In conclusion, we investigated the growth kinetics of Mn incorporation in Ga-polar GaN under different Ga/N flux conditions at a substrate temperature of 680\celsius. An approximately two-order-of-magnitude difference was observed between Ga-rich and N-rich growth regimes, with the no-flux condition exhibiting an intermediate Mn incorporation level. Although the Mn flux can influence the sticking coefficients, the Mn incorporation is expected to be governed primarily by the growth regimes. The observed flux dependence is similar to that reported for Mg dopant incorporation, for which N-rich conditions also lead to higher Mg incorporation than Ga-rich conditions.\cite{namkoong2008metal} It is also well known that Mg incorporation of more than 1~ML ($\approx1\times10^{15}$~cm$^{-2}$) on Ga-polar GaN surface results in polarity inversion.\cite{ramachandran1999inversion} For the Mn flux and growth temperature in this study, no crystal polarity inversion was observed. However, a previous report has shown that significant Mn surface accumulation can lead to polarity inversion at a relatively low growth temperature of 550\celsius.\cite{haider2003ga} In this context, future studies addressing the influence of polarity and substrate temperature on Mn incorporation would be informative, given that Mg incorporation in N-polar GaN is approximately one order of magnitude lower than in Ga-polar GaN under Ga-rich conditions and exhibits Arrhenius-type substrate temperature dependence.\cite{ptak2001magnesium,cho2017single}

\section{Acknowledgements}
\label{Ack}
YongJin~Cho thanks Oliver~Bierwagen (PDI) for reviewing the manuscript. This work was supported by the NSF FuSe project (Award 2329063), and by SUPREME, one of the seven centers in JUMP 2.0, a Semiconductor Research Corporation (SRC) program sponsored by DARPA.  

\section{Author Declarations}
\subsection{Conflict of interest}
The authors have no conflicts to disclose.

\section{Data Availability}
\label{Data}
The data that support the findings of this study are available from the corresponding authors upon reasonable request.

\bibliographystyle{aipnum4-1}
\bibliography{references}
\end{document}